\documentclass{aa}  

\newcommand\dosingle[1]{#1}  \newcommand\dodouble[1]{ } 

\newcommand\nice[1]{#1}    \newcommand\subm[1]{}   

\newcommand\monopoly[2]{#1} 

\subm{ \renewcommand\dosingle[1]{ }   \renewcommand\dodouble[1]{#1} }

\dodouble{ \documentclass[referee]{aa} }

\usepackage{graphicx}

\usepackage{amsfonts}
\usepackage{url} 

\newcommand\postrefereechanges[1]{#1}

 \usepackage{hyperref}

\nice{ 
\providecommand{\eprint}[1]{\href{http://arxiv.org/abs/#1}{{\tt [arXiv:#1]}}} 

\providecommand{\url}[1]{\href{#1}{#1}}

}

\providecommand{\adsurl}[1]{} 
\newcommand\SSS{Sect.~}
\providecommand\PRL{PRL}
\providecommand\jetpL{JETPLett}

\nice{ \newcommand\mycaptionfont{\protect\footnotesize} }

\newcommand\gtapprox{\,\lower.6ex\hbox{$\buildrel >\over \sim$} \, }
\newcommand\ltapprox{\,\lower.6ex\hbox{$\buildrel <\over \sim$} \, }
\newcommand\propapprox{\,\lower.6ex\hbox{$\buildrel \propto\over \sim$} \, }

\newcommand\arcs{\ifmmode {'' }\else $'' $\fi}     
\newcommand\arcm{\ifmmode {' }\else $' $\fi}       
\newcommand\ddeg{\ifmmode^\circ\else$^\circ$\fi}    

\newcommand\frtoday{Le\space\number\day\space\ifcase\month\or
  janvier\or f\'evrier\or mars\or avril\or mai\or juin\or
  juillet\or ao\^ut\or septembre\or octobre\or novembre\or 
d\'ecembre\fi\space \number\year}
\newcommand\todayISO{\number\year-\ifnum\month<10 0\fi\number\month-\ifnum\day<10 0\fi\number\day}

\newcommand\cqg{ClassQuantGra}   %

\newcommand\notea{^\mathrm{a}}
\newcommand\noteb{^\mathrm{b}}
\newcommand\notec{^\mathrm{c}}

\title{On the suspected timing error in Wilkinson microwave anisotropy probe map-making}

\author{Boudewijn F. Roukema 
}

\institute{Toru\'n Centre for Astronomy, Nicolaus Copernicus University,
ul. Gagarina 11, 87-100 Toru\'n, Poland 
}

\date{\frtoday}

\titlerunning{Suspected WMAP timing error}
\authorrunning{Roukema}

\begin{document}


\newcommand\Nchainsmain{16}
\newcommand\Npergroup{four}

\abstract
{It has recently been suggested that the compilation of the
  calibrated time-ordered-data (TOD) of the Wilkinson Microwave Anisotropy Probe
  (WMAP) observations of the cosmic microwave background (CMB) into
  full-year or multi-year maps may have been carried out with a small
  timing interpolation error. \protect\postrefereechanges{A large fraction} of the previously
  estimated WMAP CMB quadrupole signal would be an artefact of
  incorrect Doppler dipole subtraction if this hypothesis were
  correct.  }
{Since observations of bright foreground objects constitute part of
  the TOD, these can be used to test the hypothesis. }
{Scans of an object in different directions should be shifted by the
  would-be timing error, causing a blurring effect.  For each of
  several different timing offsets, three half-years of the
  calibrated, filtered WMAP TOD are compiled individually for the four
  W band differencing assemblies (DA's), with no masking of bright
  objects, giving 12 maps for each timing offset. Percentiles of the
  temperature-fluctuation distribution in each map at HEALPix
  resolution $N_{\mathrm{side}}=2048$ are used to determine the
  dependence of all-sky image sharpness on the timing offset. The Q
  and V bands are also considered.}
{In the W band, which is the band with the shortest exposure times,
  the 99.999\% percentile, i.e. the temperature fluctuation in the
  $\approx$ 503-rd brightest pixel, is the least noisy percentile as a
  function of timing offset. Using this statistic, the hypothesis that
  a $-25.6$~ms offset relative to the timing adopted by the WMAP
  collaboration gives a focus at least as sharp as the uncorrected
  timing is rejected at $4.6\sigma$ significance, assuming Gaussian
  errors and statistical independence between the maps of the 12
  DA/observing period combinations.  The Q and V
  band maps also reject the $-25.6$~ms offset hypothesis at high
  statistical significance. }
{The requirement that the correct choice of timing offset must
  maximise image sharpness implies that the hypothesis of a timing
  error in the WMAP collaboration's compilation of the WMAP
  calibrated, filtered TOD is rejected at high statistical
  significance in each of the Q, V and W wavebands. However, the
  hypothesis that a timing error was applied during {{\em
      calibration\/}} of the raw TOD, \protect\postrefereechanges{leading to a dipole-induced difference}
  signal, is not excluded by this method. }

\keywords{
cosmology: observations -- cosmic background radiation}

\maketitle

\dodouble{ \clearpage } 


\newcommand\tnoise{
\begin{table}
\caption{\mycaptionfont 
Percentile noisiness: mean of $\sigma_{s_p}$ (over $\delta t$) in the W band.
\label{t-noise}}
$$\begin{array}{c cc ccc} \hline
\rule[-1.5ex]{0ex}{4.5ex}
p 
& 99\%  
& 99.9\%  
& 99.99\%  
& 99.999\%  
& 99.9999\%  
\\ 
\langle \sigma_{s_p} \rangle_{\delta t} 
&  0.056
&  0.041
&  0.038
&  0.023
& 0.051
\\ \hline
\end{array}$$
\end{table}
}  

\newcommand\tQV{
\begin{table}
\caption{\mycaptionfont Mean difference in temperature-fluctuation percentiles
  $\Delta(\delta t') := \left< \; (\delta T/T)_p(\delta t = 0.5) - (\delta T/T)_p(\delta
  t') \; \right>$ for 6 DA/year combinations in each of the Q and V bands in mK, where
  $p=99.999\%$, and standard error in the mean and significance level.${\notea}$
\label{t-QV}}
$$\begin{array}{ccc  cccc } \hline
\rule[-1.5ex]{0ex}{4.5ex}
\mathrm{band} & &  
\Delta(\delta t'=0) & \Delta(0.167{\noteb}) & \Delta(0.25{\notec})
&\sigma_{\Delta} &  \Delta /\sigma_{\Delta} \\
\hline
\rule[-1.5ex]{0ex}{4.5ex} 
Q && 1.44 & -    & - & 0.05 & 29 \\
Q &&  -   &  - & 0.44 & 0.08 & 5.5 \\
V && 0.70 & -    & - & 0.06 & 11.7 \\
V && -    & 0.28 & - & 0.06 & 4.8 \\ 
\hline
\end{array}$$
\\
${}{\notea}$The time offsets $\delta t$ and $\delta t'$ are in units of one observing interval, 
i.e. 102.5~ms or 76.8~ms for Q or V, respectively. 
\\
${}{\noteb}$A shift of $-25.6$~ms relative to the WMAP collaboration choice in the V band.
\\
${}{\notec}$A shift of $-25.6$~ms relative to the WMAP collaboration choice in the Q band.
\\ 
\end{table}
}  

\newcommand\fWmean{
\begin{figure}
\centering 
\includegraphics[width=8cm]{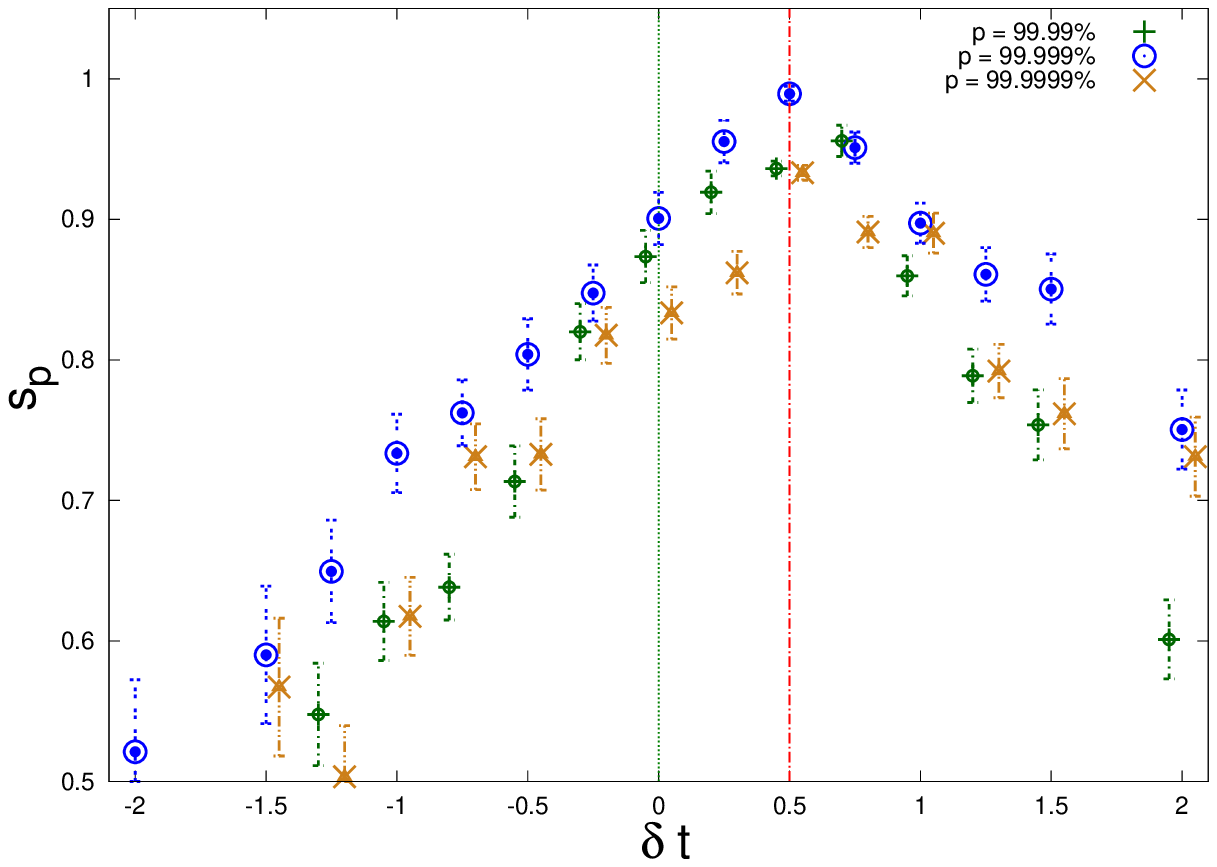}
\caption[]{ \mycaptionfont Mean normalised temperature-fluctuation
  percentiles $s_p = \left< s_p^i \right>$
  [Eq.~(\protect\ref{e-s-defn})] in compilations of W band WMAP TOD,
  for pixel fractions $p=99.99\%, 99.999\%,$ and $99.9999\%,$ as a
  function of fractional timing offset $\delta t$. The values $\delta
  t=0$ and $\delta t=0.5$, favoured by \protect\nocite{LL10toffset}{Liu} {et~al.} (2010) and
  the WMAP collaboration, respectively, are shown as vertical lines.
  The error bars are standard errors in the mean $\sigma_{s_p}$ over
  the 12 DA/year combinations. For clarity, offsets in $\delta t$ of
  $\pm 0.05$ are applied to the $p=99.99\%$ ($+$) and $99.9999\%$
  ($\times$) data points.  }
\label{f-Wmean}
\end{figure} 
} 

\newcommand\fartificialquadrupole{
\begin{figure*}
\centering 
\includegraphics[width=16cm]{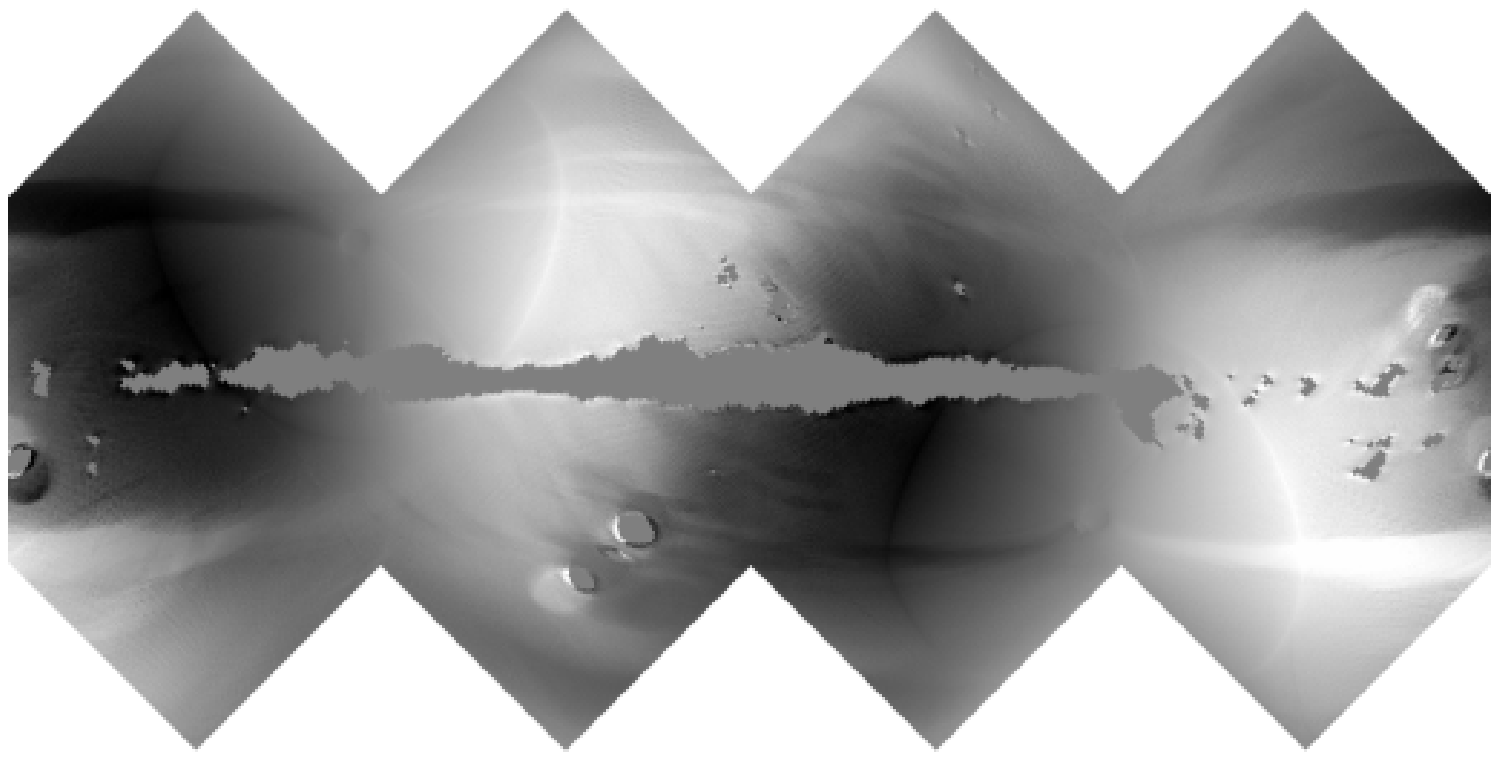}
\caption[]{ \mycaptionfont 
\protect\postrefereechanges{Dipole-induced difference} all-sky map using the WMAP quaternion history of the first year of 
WMAP observations and $\delta t=0$ in the V band, i.e.
a timing offset of $-38.4$~ms
(see \SSS\protect\ref{s-LLmapmaking}), shown in the 
hybrid, cylindrical equal-area \protect\nocite{Lambert1772}({Lambert} 1772), interrupted 
Collignon \protect\nocite{Coll1865}({Collignon} 1865) spherical projection, i.e.
the HEALPix ($H=4, K=3$) spherical projection
\protect\monopoly{\protect\nocite{RL04HPX,CR07,Gorski04HPX}({Roukema} \& {Lew} 2004; {Calabretta} \& {Roukema} 2007; {G{\'o}rski} {et~al.} 2005),}{\protect\nocite{CR07,Gorski04HPX}({Calabretta} \& {Roukema} 2007; {G{\'o}rski} {et~al.} 2005),}
centred on the Galactic Centre, with
latitude increasing upwards and longitude increasing to the left.
The grey scale is linear from $-10\mu$K (black) to $+10\mu$K (white).
No sky observations are used, i.e. only \nocite{LL10toffset}{Liu} {et~al.} (2010)'s Eqs~(1) and (2) as
encoded in their software are used to process the quaternion data. Standard
sky masks are used. This is the 80-th map iteration.
}
\label{f-yr1V1mr80}
\end{figure*} 
} 

\newcommand\fGCmfive{
\begin{figure}
\centering 
\includegraphics[width=8cm]{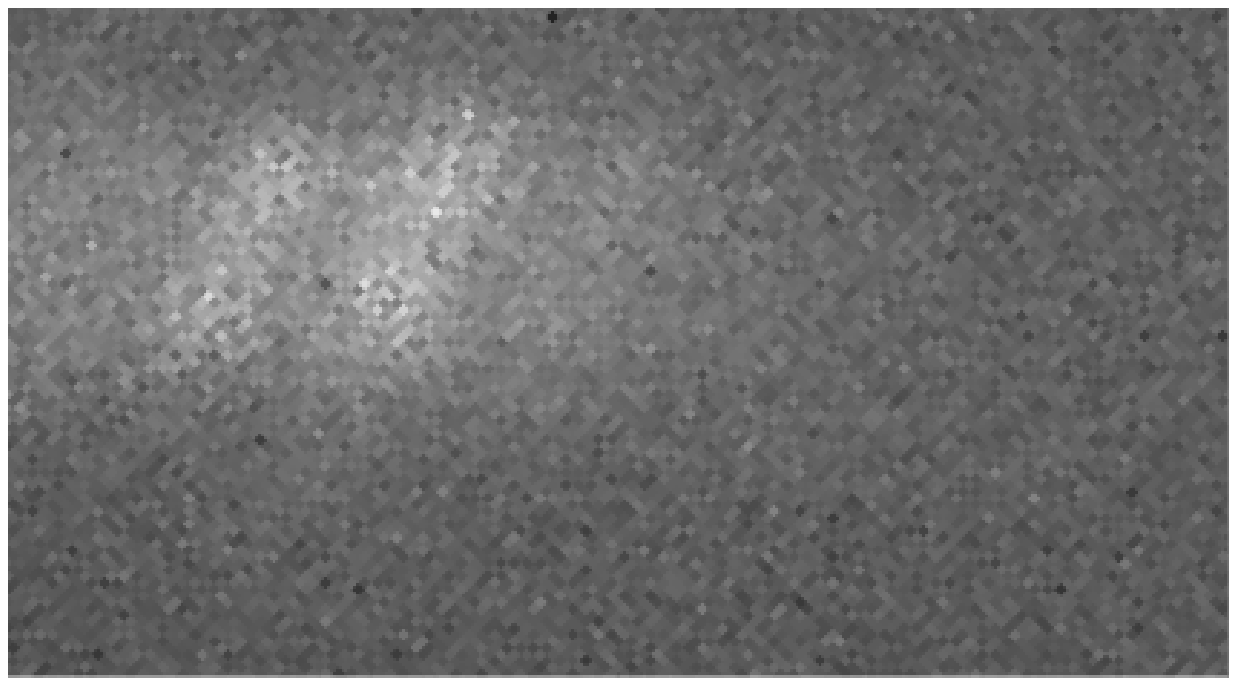}
\caption[]{ \mycaptionfont W band cylindrical equal-area projection of 
  the Galactic Centre using 199 days of the 2002 WMAP TOD
  (\SSS\protect\ref{s-obs}), compiled using a minor modification of
  \protect\nocite{LL10toffset}{Liu} {et~al.} (2010)'s analysis script
  (\SSS\protect\ref{s-LLmapmaking}), for a fractional timing offset
  $\delta_t = -5$ (i.e. an offset of $-256$~ms relative to that of
  \protect\nocite{LL10toffset}{Liu} {et~al.}). Galactic latitude increases upwards,
  galactic longitude increases to the left. The image size is
  $5.8\ddeg \times 1.7\ddeg$, centred at the Galactic
  Centre. Individual pixels are 1.7{\arcm} in side length. 
  The  temperature-fluctuation grey scale ranges from black (-20~mK) to
  white (+40~mK).  }
\label{f-GCm5}
\end{figure} 
} 

\newcommand\fGCzero{
\begin{figure}
\centering 
\includegraphics[width=8cm]{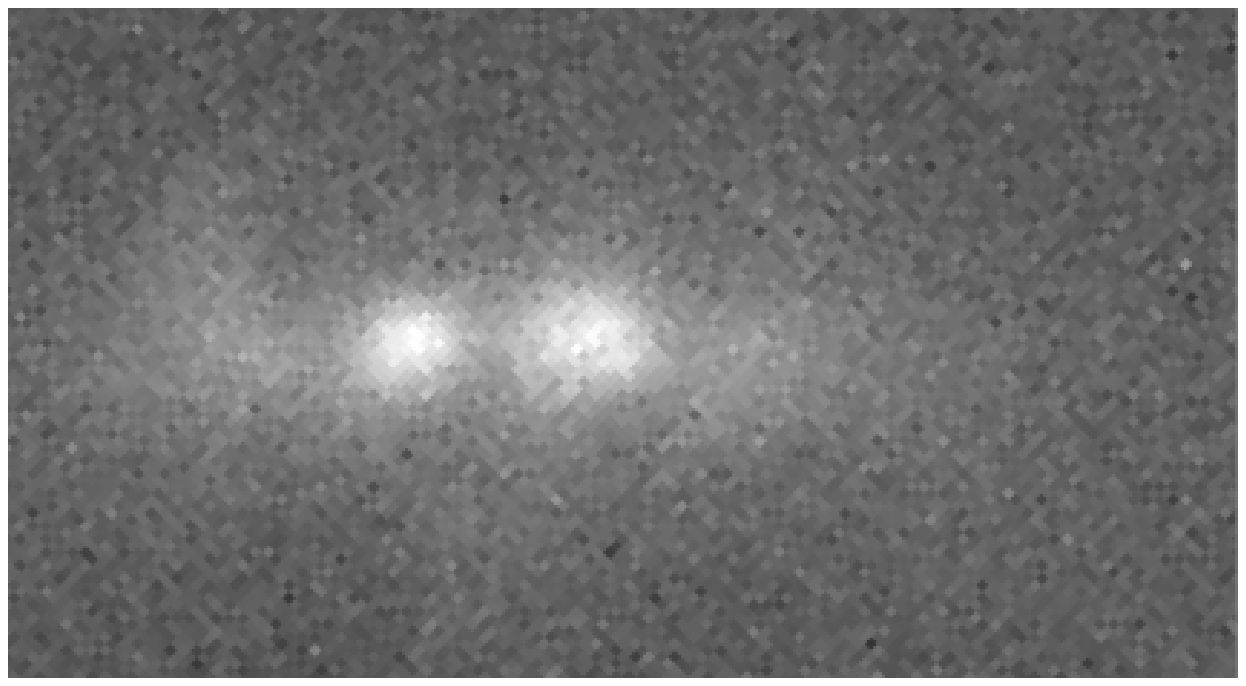}
\caption[]{ \mycaptionfont
Image of the Galactic Centre, as in Fig.~\protect\ref{f-GCm5},
for a fractional timing offset
$\delta_t = 0$ (proposed by \protect\nocite{LL10toffset}{Liu} {et~al.} 2010). 
}
\label{f-GC00}
\end{figure} 
} 

\newcommand\fGCpzerofive{
\begin{figure}
\centering 
\includegraphics[width=8cm]{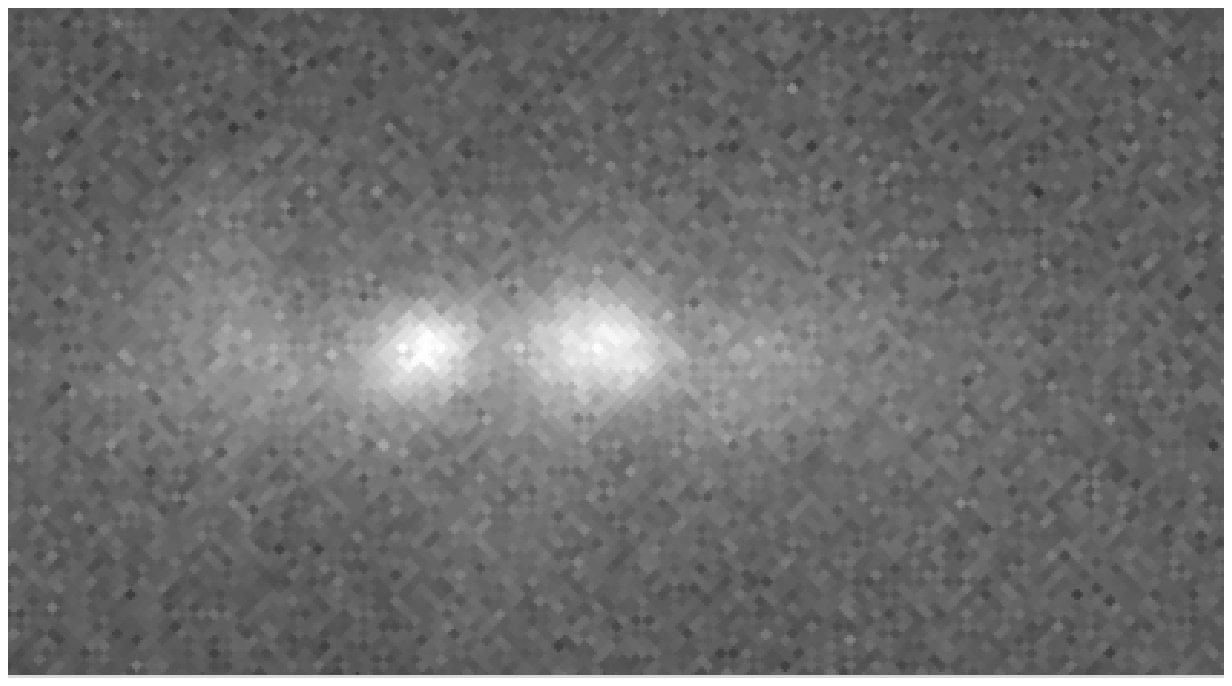}
\caption[]{ \mycaptionfont
Image of the Galactic Centre, as per Fig.~\protect\ref{f-GCm5},
for a fractional timing offset
 $\delta_t = 0.5$ (WMAP collaboration value). 
}
\label{f-GCp05}
\end{figure} 
} 

\newcommand\fGCpfive{
\begin{figure}
\centering 
\includegraphics[width=8cm]{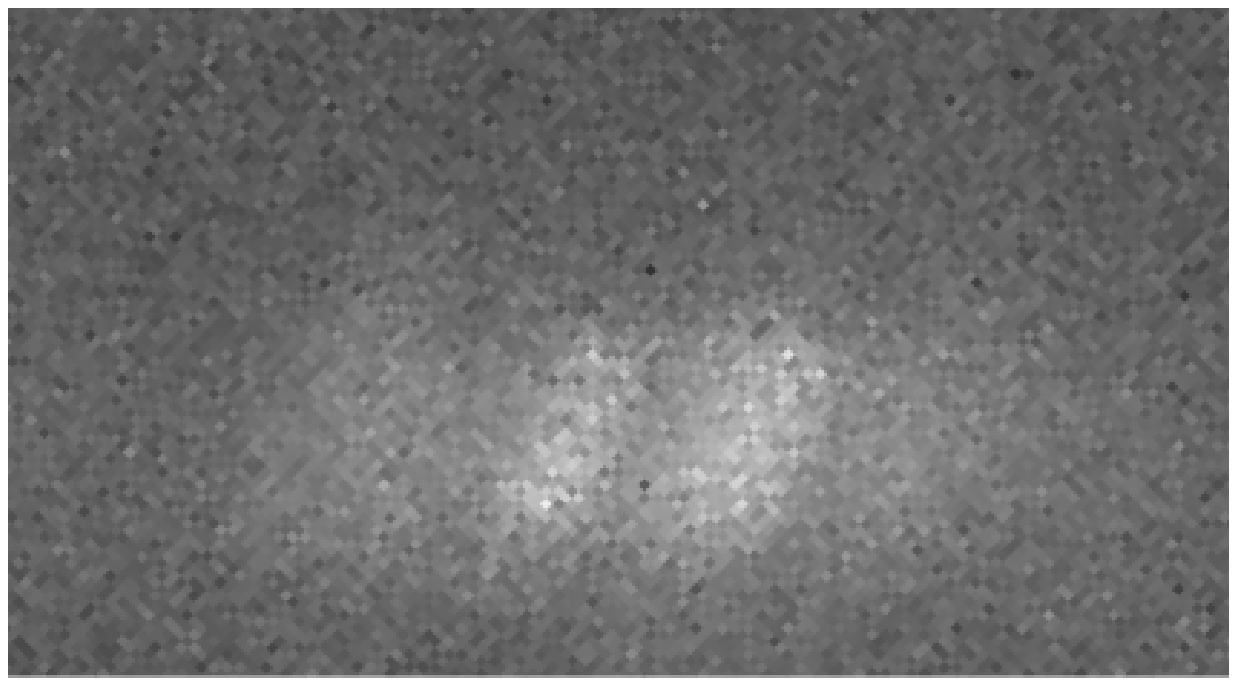}
\caption[]{ \mycaptionfont
Image of the Galactic Centre, as per Fig.~\protect\ref{f-GCm5},
for a fractional timing offset
 $\delta_t = 5$. 
}
\label{f-GCp5}
\end{figure} 
} 


\section{Introduction}  \label{s-intro}
The nature of the large-scale cosmic microwave background (CMB) signal
in the Wilkinson Microwave Anisotropy Probe (WMAP) observations
\nocite{WMAPbasic}({Bennett} {et~al.} 2003b) is of fundamental importance to observational
cosmology. A lack of structure on the largest scales in maps of CMB
temperature fluctuations would be a hint that the comoving spatial 
section of the Universe is multiply connected
\nocite{deSitt17,Fried23,Fried24,Lemaitre31ell,Rob35}({de Sitter} 1917; {Friedmann} 1923, 1924; {Lema{\^i}tre} 1931; {Robertson} 1935)
and that its size may have been detected
\nocite{Star93,Stevens93}({Starobinsky} 1993; {Stevens} {et~al.} 1993). The WMAP observations have confirmed that
the large-scale CMB signal analysed either as a quadrupole signal or
as a two-point autocorrelation function signal is weak
\nocite{WMAPSpergel,Copi07,Copi09}({Spergel} {et~al.} 2003; {Copi} {et~al.} 2007, 2009).  Several analyses indicate that
either a $T^3$ comoving space
\nocite{WMAPSpergel,Aurich07align,Aurich08a,Aurich08b,Aurich09a}({Spergel} {et~al.} 2003; {Aurich} {et~al.} 2007; {Aurich} 2008; {Aurich} {et~al.} 2008, 2010) or a
Poincar\'e dodecahedral space ($S^3/I^*$) model
\nocite{LumNat03,Aurich2005a,Aurich2005b,Gundermann2005,Caillerie07,RBSG08,RBG08}({Luminet} {et~al.} 2003; {Aurich} {et~al.} 2005a, 2005b; {Gundermann} 2005; {Caillerie} {et~al.} 2007; {Roukema} {et~al.} 2008a, 2008b)
fit the WMAP data better than an infinite, simply connected flat
model, while other analyses disagree \nocite{KeyCSS06,NJ07}({Key} {et~al.} 2007; {Niarchou} \& {Jaffe} 2007). Highly significant
evidence for either an infinite or a finite model of comoving space has not yet
been obtained.

\nocite{LL08hotsources}{Liu} \& {Li} \nocite{LL08hotsources,LL09Nobs,LL09lowquad,LL10toffset}({Liu} \& {Li} 2008; {Li} {et~al.} 2009; {Liu} \& {Li} 2009; {Liu} {et~al.} 2010) have suggested
that several systematic errors may have been made in the data analysis
pipelines used by the WMAP collaboration in the compilation of their
original time-ordered-data (TOD) into single-year or 
\postrefereechanges{multi-year} maps.
\nocite{LL08hotsources}{Liu} \& {Li} (2008) recommended that new maps should be made using
the full TOD in order to avoid some of these suspected errors.  This
requires moderately heavy computational resources (RAM and CPU) on
typical present-day desktop computers.  \nocite{Aurich09a}{Aurich} {et~al.} (2010) sidestepped
this by using a phenomenological correction for the hot pixel effect
\nocite{LL08hotsources}({Liu} \& {Li} 2008). They found that their correction applied to the
5-year WMAP W band data removes the anti-correlation in temperature
fluctuations at angular separations of nearly 180{\ddeg} and improves
the fit of a $T^3$ model to the data.  

\fartificialquadrupole 

\nocite{LL09lowquad}{Liu} \& {Li} (2009) carried out a full analysis of the 5-year WMAP
TOD.  
\postrefereechanges{Their analysis pipeline appeared to give test 
results largely compatible with those of the
WMAP collaboration.  However, the CMB quadrupole amplitude was found to be
much weaker, and sub-degree power was also found to be a little weaker.}
Later, \nocite{LL10toffset}{Liu} {et~al.} (2010) traced the difference between their analysis
and that of the WMAP collaboration to a difference in interpolating
the timing of individual observations from the times recorded in the
TOD files. The TOD FITS format files contain observational
starting times in both the ``Meta Data Table'' and the ``Science Data
Table''. 
The full set of observing times of individual
observations are not recorded in either data table; they must be
interpolated from the smaller set of observing times that are recorded.  
The starting times in the Meta Data Table are 25.6~ms greater
than the corresponding times in the Science Data Table. 
A computer script for checking this is given in Appendix~\ref{a-meta_vs_science}.
The durations of
individual observations in the Q, V and W bands are 102.4~ms, 76.8~ms,
and 51.2~ms, respectively \nocite{WMAPExplSupp100405}(Section 3.2,
 {Limon} {et~al.} 2010). 
The 21 April 2010 version of the WMAP Explanatory Supplement 
\nocite{WMAPExplSupp100405}(Section 3.1,  {Limon} {et~al.} 2010) discusses the relation between the start times
in the two tables, but does not refer to the 25.6~ms offset, i.e. half
of a W band observing interval or a quarter of a Q band observing
interval. \postrefereechanges{\nocite{Bennett03MAP}{Bennett} {et~al.} (2003a) state (\SSS{6.1.2}) that 
``a relative accuracy of 1.7~ms can be achieved between the star tracker(s), gyro and
the instrument'', i.e. the combined random and systematic error 
should be much smaller than 25.6~ms.}
Given that the timing offset is numerically implicit in the WMAP TOD files, it could, 
strictly speaking, be referred to as the WMAP collaboration's ``implicitly claimed but ignored''
timing offset. However, for simplicity, the timing offset will be referred to here as \nocite{LL10toffset}{Liu} {et~al.} (2010)'s
hypothesis.

Using their Eqs~(1) and (2), \nocite{LL10toffset}{Liu} {et~al.} (2010) showed that the error
in estimating the Doppler dipole of the spacecraft velocity implied by
the would-be timing error can be used to produce a map \nocite{LL10toffset}(Fig.~2
  left,  {Liu} {et~al.} 2010) that matches the direction
and amplitude of the CMB quadrupole signal as estimated by the WMAP
collaboration \nocite{LL10toffset}(Fig.~2 right,  {Liu} {et~al.} 2010) remarkably well, despite using
only the history of spacecraft attitude quaternions (representing
three-dimensional directions) and no sky observations. 
This is easy to check using the WMAP calibrated, filtered, 3-year
TOD and \nocite{LL10toffset}{Liu} {et~al.} (2010)'s software. Figure~\ref{f-yr1V1mr80} shows 
a \protect\postrefereechanges{dipole-induced difference} 
map\footnote{\protect\postrefereechanges{\nocite{LL10toffset}{Liu} {et~al.} (2010) refer to this
as a ``differential dipole'' map. However, this term is frequently used
to refer to monopole-subtracted dipole maps. 
Moreover, the difference between two
(fixed) dipole signals in different directions is itself also a dipole, not 
a quadrupole. To avoid misinterpretations, the term ``dipole-induced 
difference'' is suggested here.}}
 which is consistent with \nocite{LL10toffset}{Liu} {et~al.} (2010)'s \protect\postrefereechanges{dipole-induced difference} map
shown in their Fig.~2 (left), 
for a half-observing-interval offset in the V band, i.e. $-38.4$~ms.
Using a velocity dipole model and a $-25.6$~ms
offset, \nocite{MSS10}{Moss} {et~al.} (2010) find a similar 
\postrefereechanges{but weaker effect of reduction in the quadrupole amplitude.}
\postrefereechanges{Earlier, \nocite{Jarosik06Beam}{Jarosik} {et~al.} (2007) (\SSS{2.4.1}) noted that 
an error during the calibration process could introduce an artificial,
dipole-induced quadrupole.}

Interpolation of the observing times and sky positions from the WMAP
TOD files cannot be avoided. Given that the WMAP Explanatory Supplement
does not presently account for the 25.6~ms offset between the two data tables in
each TOD file, and given the potential importance of a 70--80\%
overestimate of the CMB quadrupole amplitude, it is useful to see
if a relatively simple, robust method of analysing the TOD can
determine which of the two timing interpolation methods, that of the
WMAP collaboration or that proposed by \nocite{LL10toffset}{Liu} {et~al.} (2010), is
correct. 

The WMAP TOD include many observations of bright foreground objects,
including Solar System planets and objects in the Galactic Plane.  The
complex scanning pattern of WMAP implies that individual objects are
likely to be scanned in different directions over a series of
successive observations.  The greatest differences in scanning
directions are likely to occur over observing intervals of weeks or
more. An error in the assumed sky direction due to a timing error should
cause observations of a compact source made at different times to be
slightly offset from one another in the compiled map, i.e. it should
cause an overall shift in sky position and a blurring effect. For a
large enough timing offset, double imaging of compact sources should occur
(public communication, Crawford 2010\footnote{\url{http://cosmocoffee.info/viewtopic.php?t=1537}};
see also \nocite{MSS10}{Moss} {et~al.} 2010). 
The W band
observations, with the highest angular resolution, have an estimated
FWHM beam size of about 12--12.6{\arcm} \nocite{WMAP1beam}(Table 5,  {Page} {et~al.} 2003). An error of
25.6~ms corresponds\footnote{One spin during 2.2 minutes, i.e. 
\protect\postrefereechanges{129.3~s, \protect\nocite{WMAP03syserr}(e.g. Sect~3.4.1,  {Hinshaw} {et~al.} 2003)}
corresponds to $360\ddeg \sin(70.5\ddeg) = 339\ddeg$ in great circle degrees, so 25.6~ms corresponds
to \postrefereechanges{4.0{\arcm}}. } to an error of about \postrefereechanges{4.0{\arcm}}. 
Hence, a slight blurring of the images of compact sources rather than
double imaging should occur. 
\postrefereechanges{If the WMAP collaboration's timing offset were correct, 
then this blurring could explain the slight loss of 
sub-degree power found by \nocite{LL10toffset}{Liu} {et~al.} (2010) as an artefact for the latter's choice of timing
offset.} \postrefereechanges{The blurring} could be detectable in beam shape
analysis, but this may not be easy.  \nocite{WMAP1beam}{Page} {et~al.} (2003) model the beam
shape based on the physical structure of the WMAP receivers, but 
estimate centroids iteratively.
It is not obvious that this iterative procedure would have detected an
artificial blurring effect of just a few arcminutes. 

\fGCmfive

\fGCzero
\fGCpzerofive

\fGCpfive

Nevertheless, the blurring effect should be statistically detectable
in maps compiled from the TOD. 
The effect should be strong if the timing error is exaggerated beyond that
proposed by \nocite{LL10toffset}{Liu} {et~al.} (2010). By considering the timing offset as a free parameter,
it should be possible to find the value that minimises blurring. This should 
correspond to the correct choice of timing interpolation.
In principle, 
since a timing error also induces an error in absolute sky position,
this could also be used to determine the correct method. 
However, it is conceivable that a positional offset may have
been at least partially removed at some stage in the pointing calibration, making it hard
to detect, as suggested by \nocite{MSS10}{Moss} {et~al.} (2010). 
On the other hand, a blurring effect is unlikely to have
been removed at any step in the production of WMAP maps. Hence, the analysis
here uses the blurring effect. 
In \SSS\ref{s-method}, an empirical method of estimating the blurring
using all-sky maps is presented. Images of some Galactic Centre objects as a function of
timing offset and the results of the statistical analysis are presented in 
\SSS\ref{s-results}. Conclusions are given in \SSS\ref{s-conclu}.

\section{Method} \label{s-method}

\subsection{Observational data files and differencing assemblies} \label{s-obs}
Compilation of TOD into maps requires a moderately heavy usage of RAM,
CPU and disk space on present-day desktop computers. For this reason,
only a subset of the available WMAP TOD is used rather than the full
TOD. From each of the years 2002, 2003 and 2004, the files for the
first (by filename) 199, 199, and 198 days,\footnote{The year 2004 has only 198 days because 
the file that would have the label ``20041482358\_20041492358''
is absent from the data set.} respectively, of filtered, calibrated
TOD\footnote{\href{http://lambda.gsfc.nasa.gov/product/map/dr2/tod_fcal_get.cfm}{{\tt http://lambda.gsfc.nasa.gov/product/map/dr2/}}
\href{http://lambda.gsfc.nasa.gov/product/map/dr2/tod_fcal_get.cfm}{{\tt tod\_fcal\_get.cfm}}}
are used, i.e. slightly over six months of data during each year. This allows
a convenient filename \postrefereechanges{pattern}.
\postrefereechanges{\nocite{LL09lowquad}{Liu} \& {Li} (2009) analysed the unfiltered,
  calibrated, 5-year WMAP data release. The filtering should not
  significantly affect the questions of interest here.}

The primary analysis is done here using the W band, since this has the best
resolution. There are four W differencing assemblies (DA's), giving 12 TOD
sets that in many ways can be considered to constitute independent observations.
Complementary analyses using the Q and V DA's, with 6 TOD sets each, are
also carried out.

\subsection{Compilation of the TOD}    \label{s-LLmapmaking}
\nocite{LL10toffset}{Liu} {et~al.} (2010) have made their script for compiling the TOD 
publicly available.\footnote{\url{http://cosmocoffee.info/viewtopic.php?p=4525}, 
\href{http://dpc.aire.org.cn/data/wmap/09072731/release_v1/source_code/v1/}{{\tt http://dpc.aire.org.cn/data/wmap/09072731/release\_v1/}} 
\href{http://dpc.aire.org.cn/data/wmap/09072731/release_v1/source_code/v1/}{{\tt source\_code/v1/}}}
For the purposes of the present analysis, 
minor modifications of this script and associated libraries have been made in order to run
the script using the GNU Data Language 
(GDL).\footnote{\href{http://cosmo.torun.pl/GPLdownload/LLmapmaking_GDLpatches/LLmapmaking_GDLpatches_0.0.3.tbz}{{\tt http://cosmo.torun.pl/GPLdownload/}} 
\href{http://cosmo.torun.pl/GPLdownload/LLmapmaking_GDLpatches/LLmapmaking_GDLpatches_0.0.3.tbz}{{\tt LLmapmaking\_GDLpatches/LLmapmaking\_GDLpatches\_0.0.3.tbz}}}
In particular, the data flag parameter {\sc daf\_mask} 
has been set to 1 in order to include all planet observations when compiling the TOD, and
the processing mask has been set to 0 in order not to hide any bright objects from
the analysis. 

Moreover, the {\sc center} parameter for determing the timing offset
is generalised to an arbitrary floating point value. Here, this is written
$\delta t$, i.e. the timing offset expressed as a fraction of 
an observing time interval in a given band. This is used as a fraction for
interpolation between quaternions. Thus, $\delta t = 0$ is 
\nocite{LL10toffset}{Liu} {et~al.} (2010)'s primary hypothesis of what is the correct offset, and $\delta t = 0.5$ is what
the WMAP collaboration believes to be correct. In the W band, $\delta t = 0$ corresponds
to $-25.6$~ms relative to $\delta t=0.5$.
In order to reduce computational time, map iteration is not done. 

\postrefereechanges{The planets are useful for this analysis because
  of their high brightness.  However, some of them (especially
  Jupiter) move by up to a few degrees during the times when they
  happen to be scanned. In directions orthogonal to their movement, a
  blurring effect is to be expected for a wrong choice of timing
  offset. In directions parallel to their movement, a wrong choice of
  timing offset can either cause blurring, or cause some observations
  that should be mapped to different pixels to incorrectly be mapped
  to the same pixel, i.e. creating a false focussing effect. The
  degree to which the latter effect is important depends on the degree
  to which the observations of a given planet are discontinuous and
  the way that this discretisation interacts with the scanning pattern
  directions. In practice, semi-quantitative examination of images of
  Jupiter suggests that the false focussing effect is present, but
  weak, for some values of $\delta t$. An analysis using
  Jupiter alone would need to take this effect into account.}

\subsection{Pixel resolution and the blurring effect}   \label{s-delta_t}
Since the blurring effect expected from a 25.6~ms timing error would be on 
a scale of \postrefereechanges{$\sim 4.0${\arcm}},
the commonly used HEALPix resolution 
\nocite{Healpix98}({G\'orski} {et~al.} 1999) of $N_{\mathrm{side}}=512$ would be insufficient, since this gives pixel
sizes of 6.9{\arcm}. However, very high resolutions would give very large 
file sizes. 
As a compromise, $N_{\mathrm{side}}=2048$ is adopted here, giving a pixel size of 1.7{\arcm} 
and one-dimensional FITS binary
table file sizes of about 190~Mib, or two-dimensional FITS image file sizes
of about 400~Mib, using the 
the HEALPix ($H=4, K=3$) spherical projection 
\protect\monopoly{\nocite{RL04HPX,CR07,Gorski04HPX}({Roukema} \& {Lew} 2004; {Calabretta} \& {Roukema} 2007; {G{\'o}rski} {et~al.} 2005).}{\nocite{CR07,Gorski04HPX}({Calabretta} \& {Roukema} 2007; {G{\'o}rski} {et~al.} 2005).}
The utility program {\sc HPXcvt}
from the {\sc WCSLIB} library can be used to convert from one-dimensional to 
two-dimensional formats.\footnote{\url{http://www.atnf.csiro.au/people/mcalabre/WCS/}}
Image files that use the HEALPix spherical projection can be viewed
using standard astronomical FITS image viewing tools for visual
checking of the blurring effect.

Figures~\ref{f-GCm5}--\ref{f-GCp5} show the Galactic Centre from the
2002 WMAP TOD compiled as discussed above, for $\delta_t = -5, 0,
0.5,$ and $5$.  Figures~\ref{f-GC00} and \ref{f-GCp05} show the 
timing options of \nocite{LL10toffset}{Liu} {et~al.} (2010) and the WMAP collaboration,
respectively. 
The blurring effect is not obvious by visual inspection. These
two images appear about equally sharp.
In contrast, Figs~\ref{f-GCm5} and \ref{f-GCp5}, in which the timing offsets are
greater, clearly show both a positional offset and blurring.

\subsection{Temperature-fluctuation distribution percentiles} \label{s-percentiles}
The blurring in Figs~\ref{f-GCm5} and \ref{f-GCp5} is the key to the method used here.
The numbers of very
bright pixels in these two figures are much lower than in Figs~\ref{f-GC00} and
\ref{f-GCp05}.  This suggests a statistic to characterise the degree
of sharpness of the image: the number of pixels $N_{\mathrm{b}}(\delta
T/T, \delta t)$ above a given temperature fluctuation threshold
$\delta T/T$ in a map calculated at a given timing offset $\delta t$. Maximising
$N_{\mathrm{b}}$ over $\delta t$ should give the map most likely to be
correct.

The brighter the threshold $\delta T/T$, the fewer
the number of pixels that will lie above this threshold, increasing
Poisson noise. 
Thus, fainter thresholds $\delta T/T$ should give stronger signals.

The situation for intermediate brightness pixels is less simple. The
signal in the very bright pixels in Figs~\ref{f-GC00} and \ref{f-GCp05} has been
spread out into intermediate brightness pixels in Figs~\ref{f-GCm5}
and \ref{f-GCp5}, increasing the numbers of the latter.  However, the signal in regions of
intermediate brightness in Figs~\ref{f-GC00} and \ref{f-GCp05} 
has also been blurred {\em out} of the same brightness interval, down
to fainter levels.
For typical profiles of astronomical objects, 
the number of pixels in each different brightness range should increase as the
threshold decreases while remaining well above zero. Hence, the loss of
pixels that become blurred below a given threshold $\delta T/T$ should
outweigh the gain in pixels from intrinsically higher brightnesses.
That is, in general, an incorrect value of the fractional timing offset $\delta t$
should decrease $N_{\mathrm b}$ at any positive value of $\delta T/T \gg 0$.

\tnoise

\fWmean

Approaching low (positive) temperature-fluctuation levels, the measurement noise 
in $\delta T/T$
should proportionally increase, so that noise in $N_{\mathrm b}$ increases.
The effects of smoothing with negative fluctuations will also arise at
low positive temperature-fluctuation levels, reducing the effectiveness of testing
$N_{\mathrm b}$.
Hence, the least noisy threshold $\delta T /T$ for
optimising the sharpness of images in the map
should lie somewhere between the extremes of low positive $\delta T/T$ and 
high $\delta T/T$.

Rather than choosing fixed values of $\delta T/T$ and studying
$N_b(\delta T/T, \delta t)$, the approach adopted here is to consider
$\delta T/T(p N, \delta t)$, where $p$ is a fraction close to unity
and $N = 12(2048)^2$ is the number of pixels. That is, the 
percentiles of the (sorted)
temperature-fluctuation distribution are determined for several values of $p$,
as a function of the fractional timing offset $\delta t$.

Use of different DA's and 
different (half-)years of data should give estimates of the random error at any
given $p$ and $\delta t$. Averaging over $\delta t$ at a given $p$
makes it possible to find the fraction $p$ that gives the least noisy percentile.
Thus, at a given DA and year combination $i$
and a percentage $p \ltapprox 1$, the percentile is a function 
\begin{equation}
  \left( \frac{\delta T}{T} \right)_p^i \, (\delta t).
\label{e-defn-percentile}
\end{equation}
In order to give approximately equal weight to the different
samples $i$, the percentiles are normalised over $\delta t$ internally within each sample $i$ at each
$p$.
That is, $\left( {\delta T}/{T} \right)_p^i$ is evaluated on a series of maps calculated for $-5 \le \delta t \le 5$,
attaining a minimum $a_p^i$ and a maximum $b_p^i$. This is normalised to
\begin{equation}
  s_p^i(\delta t) := \frac{ \left( {\delta T}/{T} \right)_p^i (\delta t) - a_p^i}{b_p^i-a_p^i}
\label{e-s-defn}
\end{equation}
The mean and standard error in the mean of $s_p^i$ over the DA/year
combinations $i$ can then be estimated at each $\delta t$. 
These are written $s_p$ and $\sigma_{s_p}$, respectively. 

\section{Results} \label{s-results}

\subsection{W band}
Calculations on 4-core, 2.4~GHz, 64-bit processors with 4~Gib RAM,
using GDL-0.9$\sim$rc4 running under GNU/Linux, took about 3 hours per
map.  Table~\ref{t-noise} lists the mean of $\sigma_{s_p}(\delta t)$
at several values of $p$ for the W band analysis.  This shows that
$p=99.999\%$, i.e. the fraction $p$ which selects the $\approx$ 503-rd brightest pixel
in any given map, gives the least noisy percentile.
Figure~\ref{f-Wmean} shows the mean normalised temperature-fluctuation 
percentiles $s_p$ for this and two other values of $p$.
The $s_{99.999\%}$ threshold shows
a sharp maximum at $\delta t=0.5$. The higher percentile, i.e. at $p=99.9999\%$,
is noisier, but still significantly favours $\delta t = 0.5$ over any
other value of $\delta t$ evaluated. The $p=99.99\%$ percentile is also noisier
than that at $p=99.999\%$, and marginally favours $\delta t = 0.75$ 
over $\delta t = 0.5$. \nocite{LL10toffset}{Liu} {et~al.} (2010)'s hypothesis of $\delta t = 0$ 
is not favoured at any fraction $p$.

Exact symmetry around the correct value of $\delta t$ is not 
necessarily expected, since the scan paths are curved (non-geodesic) paths on the 2-sphere,
not straight lines in the 2-plane, and the source objects are not isolated from
one another. However, approximate symmetry can reasonably be expected. It is obvious that Fig.~\ref{f-Wmean}
is not symmetric around $\delta t=0$. Hiding the part of the figure to the left
of $\delta t=-1$ shows that the figure {\em does} appear to be approximately 
symmetric around $\delta t=0.5$. 

\tQV

A quantitative estimate of a lower limit to the significance to which
the $\delta t=0$ hypothesis can be rejected, given that it must
maximise $s_p(\delta t)$, can be obtained by testing the hypothesis
that $s_p^i(\delta t = 0) \ge s_p^i(\delta t = 0.5)$ under the
assumption that the error distributions are Gaussian. That is, the
hypothesis that that a $-25.6$~ms offset relative to the timing
adopted by the WMAP collaboration gives a focus at least as sharp as
that obtained using the latter's timing choice is tested.  The
percentiles shown in Fig.~\ref{f-Wmean} are $s_p(0)= 0.901 \pm 0.019$
and $s_p(0.5)= 0.989 \pm 0.005$, where $p=99.999\%$, i.e. where $s_p$
is the percentile that gives the least noisy result, so
the hypothesis is rejected at 4.6$\sigma$ significance.

\subsection{Q and V bands}
A small number of maps were calculated in the Q and V bands, which
have lower angular resolution and fewer numbers of DA's.  Only a few
maps were calculated, so the percentiles are not normalised.  Instead,
their differences are calculated.  The \nocite{LL10toffset}{Liu} {et~al.} (2010) hypothesis
may be interpreted in the Q and V bands either as a fractional time
offset $\delta t=0$ (where $\delta t=0.5$ is the WMAP collaboration's
choice), or as a fixed offset in time units of $-25.6$~ms, since this
is the undocumented difference between the start times in the Meta
Data Set and the Science Data Set in the TOD files.  The latter
difference should be more difficult to test than the former, since the
interval is smaller in both cases. These differences are calculated
individually in each DA/year combination.  Table~\ref{t-QV} shows
their statistics.  The weaker form of \nocite{LL10toffset}{Liu} {et~al.} (2010)'s hypothesis
is rejected at very high significance ($\ge 4.8\sigma$) in all cases. In other words,
the temperature fluctuation in the $\approx $503-rd brightest pixel is
lower for \nocite{LL10toffset}{Liu} {et~al.} (2010)'s hypothesised offset compared to its
value for the WMAP collaboration's hypothesised offset at very high
significance in all cases.

\section{Conclusion} \label{s-conclu}

The difference in image sharpness for \nocite{LL10toffset}{Liu} {et~al.} (2010)'s and the WMAP
collaboration's respective timing interpolation choices are not
obvious in the images of the Galactic Centre shown in
Figs~\ref{f-GC00} and \ref{f-GCp05}. However, use of the all-sky
images gives estimates of image sharpness that depend strongly on 
the fractional time offset $\delta t$.  These calculations do not
require any fitting of source or beam profiles, nor any selection of one or more
preferred sources, nor any assumptions regarding
pointing accuracy.   The data analysis pipeline is performed using
\nocite{LL10toffset}{Liu} {et~al.} (2010)'s script, with only minor
modifications, on slightly under a quarter of the full set of WMAP calibrated, 
filtered TOD. \nocite{LL10toffset}{Liu} {et~al.} (2010)'s
timing error hypothesis is rejected at very high significance in the Q, V and
W wavebands separately.

The lack of an explanation in the WMAP Explanatory Supplement
\nocite{WMAPExplSupp100405}(Section 3.1,  {Limon} {et~al.} 2010) for the 25.6~ms offset
between the start times in the Meta Data Set and the Science
Data Set in the TOD files is an unfortunate gap in the
documentation, but it clearly did not lead to an error of this
magnitude in the WMAP collaboration's compilation of the calibrated, filtered TOD.  A
possible explanation for the offset could be that it is a convention
inserted by software into the Meta Data Set that was forgotten about,
rather than a physical time offset. 
In principle, it would be good if the WMAP collaboration could 
find the line(s) in their software where the 25.6~ms offset is inserted, 
in order to confirm that its origin is fully understood, and that it
has no unintended consequences.

Nevertheless, the fact that \nocite{LL09lowquad}{Liu} \& {Li} (2009) accidentally applied a time offset
that approximately cancels the $\delta t = 0.5$ CMB quadrupole in
amplitude and direction, and that this can be modelled as a 
\protect\postrefereechanges{Doppler-dipole-induced difference} signal \nocite{LL10toffset}(Eq.~(2), {Liu} {et~al.} 2010, ``deviation of differential dipole''), remain
interesting coincidences. The relations between the
quadrupole, octupole and ecliptic \nocite{Copi06ecliptic,Copi10}(e.g.,  {Copi} {et~al.} 2006; {Sarkar} {et~al.} 2010) have
long been known. What remains unestablished is whether these are 
coincidences or artefacts.
\nocite{MSS10}{Moss} {et~al.} (2010) discuss the possibility of relationships between the
various coincidences and suggest that another 
effect might couple with the WMAP spacecraft velocity
dipole in order to produce an artificial component of the quadrupole.

One interesting possibility for further work would be to consider a
variation on the hypothesis proposed by \nocite{LL10toffset}{Liu} {et~al.} (2010): could it
be possible that the would-be timing error was introduced when {\em
  calibrating the uncalibrated\/} TOD? 
\postrefereechanges{Although the velocity dipole is only removed from
the data during the mapmaking step, it is first used for calibrating the raw
data into a calibrated signal in mK. So, a} small timing error during the calibration
step could correspond to artificially adding a \protect\postrefereechanges{dipole-induced difference}
signal.  \postrefereechanges{During the following step of compiling the calibrated TOD
into maps, the dipole would then be subtracted at the correct positions, but without
compensating for the incorrect calibration.}
\postrefereechanges{As noted above, \nocite{Jarosik06Beam}{Jarosik} {et~al.} (2007) (\SSS{2.4.1}) have 
earlier considered the possibility that a calibration error 
(not necessarily
induced by a timing error) could introduce an artificial,
dipole-induced quadrupole.}

The calibration method for the 7-year WMAP analyses are
stated \nocite{WMAP7JarosikBasic}({Jarosik} {et~al.} 2010) to be those used for the 5-year
analyses, i.e. as described in \SSS{4} of \nocite{WMAP5Hinshaw}{Hinshaw} {et~al.} (2009),
retaining the estimate of about 0.2\% absolute calibration error.
Since the dipole amplitude is about 3~mK, a 0.2\% error corresponds to
about 6~$\mu$K. The 7-year quadrupole estimated for $\delta t =0.5$ is
\postrefereechanges{$\sqrt{(3/\pi)C_2} \approx 14 \mu$K} \nocite{WMAP7JarosikBasic}(Section 4.1.1,
 {Jarosik} {et~al.} 2010), only a factor of two higher than this
calibration uncertainty (though the model-dependent systematic
error is high for the infinite flat model).
\nocite{LL09lowquad}{Liu} \& {Li} (\SSS{4}, Fig.~6,
2009) estimate the quadrupole-like difference between their Q1 DA
$\delta t=0$ map and the WMAP collaboration's $\delta t=0.5$ map to
have an r.m.s. of 6.6~$\mu$K.  Moreover, the spacecraft's velocity is
assumed by \nocite{WMAP5Hinshaw}{Hinshaw} {et~al.} (2009) and \nocite{WMAP7JarosikBasic}{Jarosik} {et~al.} (2010)
to be known exactly, and fits are made over intervals
``typically between 1 and 24 hours''.  It is not obvious that a timing
error causing the spacecraft's velocity vector to be slightly
misestimated, which in turn causes a quadrupole-like 
\protect\postrefereechanges{dipole-induced difference} signal over a year's observations, of amplitude $\sim 5$--$10
\mu$K, would have been detected or removed during the calibration
process. \postrefereechanges{Indeed, \nocite{KCover09}{Cover} (2009) claims the presence
of considerable systematic uncertainty in the calibration of the WMAP TOD.}

The calibration only applies to (differential) radio flux density estimates, not to
positional data, so it would not affect \postrefereechanges{the latter}, i.e.
the spacecraft attitude quaternions.
The Meta Data Tables containing the quaternions
 appear to be identical between the
``uncalibrated'' and ``calibrated, filtered'' WMAP 3-year TOD files, as can
be expected.  Hence, an incorrect calibration that 
\protect\postrefereechanges{leads to a dipole-induced difference} signal would not cause an arcminute-scale
blurring effect nor a positional error in point sources. It would 
not be detectable by the method used in this paper. Since a large
fraction of the brightest point sources at WMAP frequencies are
variable, it may not be easy to use point sources to calibrate the
uncalibrated data independently of the dipole calibration.
\nocite{LL09lowquad}{Liu} \& {Li} (2009) and \nocite{LL10toffset}{Liu} {et~al.} (2010) appear to use the word ``raw'' to
refer to the calibrated (filtered for 3-year analyses, unfiltered for
5-year analyses) TOD and do not appear to have carried out the calibration
step. 

Independently of these considerations, the correct way to compile the
{\em calibrated\/} WMAP TOD into sky maps must clearly be the one that
optimises the sharpness of the images of bright sources. Whether or
not the calibration step itself was carried out with a small timing
error---\protect\postrefereechanges{leading to an artificial dipole-induced difference}
signal---remains to be investigated. \nocite{LL10toffset}{Liu} {et~al.} (2010)'s
calculations suggest that the latter is an interesting possibility.

\begin{acknowledgements}
Thank you to Douglas Applegate for bringing this topic to the
attention of the cosmocoffee.info
forum (\url{http://cosmocoffee.info/viewtopic.php?t=1537}), to
Hao Liu and Ti-Pei Li for useful public and private discussion and
making their software publicly available, to Tom Crawford for
pointing out the role of scanning directions, \postrefereechanges{
to an anonymous referee for several helpful comments, and to Bartosz
Lew for useful discussion}.
Use was made of the WMAP data
(\url{http://lambda.gsfc.nasa.gov/product/}), 
the {\sc GNU Data Language}
(\url{http://gnudatalanguage.sourceforge.net/}), 
a GPL version of the {\sc HEALPix} software \nocite{Healpix98}({G\'orski} {et~al.} 1999),
the {\sc IDL Astro} distribution,
the Centre de Donn\'ees astronomiques de Strasbourg 
(\url{http://cdsads.u-strasbg.fr}),
and
the GNU {\sc Octave} command-line, high-level numerical computation software 
(\url{http://www.gnu.org/software/octave}).

%
%
%

\end{acknowledgements}

\subm{ \clearpage }

\nice{
%

}


\appendix
\section{Time offset between Meta Data Table and Science Data Table}
\label{a-meta_vs_science}

Since the time offset between the Meta and Science Data Tables in the WMAP TOD
has escaped attention for many years, a method of verifying this is provided here.

The TOD files, the GNU Data Language (GDL), and the IDL Astro library 
are available at \url{http://lambda.gsfc.nasa.gov/product/map/dr2/} 
\url{tod_fcal_get.cfm}, \\  
\url{http://idlastro.gsfc.nasa.gov/ftp/astron.dir.tar.gz}, 
\url{http://gnudatalanguage.sourceforge.net/}, respectively.
The file path for the IDL Astro library can be set up using the equivalent of
\begin{verbatim}
 cd pro && ln -s */*.pro . && export GDL_PATH=`pwd`
\end{verbatim}
in a bash shell.
The line
\begin{verbatim}
ndata = product( axis, /integer )
\end{verbatim}
in the IDL Astro routine {\sc fits/fits\_read.pro} should be replaced by
\begin{verbatim}
      ndata = floor( product( axis ) + 0.5)
\end{verbatim}
for use in GDL-0.9$\sim$rc3 or earlier. 
The following GDL script demonstrates the timing offset
in the TOD, where ``tod.fits'' is the name of a TOD file. 
\begin{verbatim}
pro TOD_offset
  fits_open, 'tod.fits', handle
  ;; get table of times (metatime) of quaternions 
  ;; of spacecraft attitude
  fits_read, handle, metatable, metaheader, $
             extname = 'Meta Data Table'
  tbinfo, metaheader, meta_h
  metatime= tbget(meta_h,metatable,'Time')

  ;; get table of times (scitime) stored in table 
  ;; that includes observational data
  fits_read,handle,scitable,sciheader, $
            extname = 'Science Data Table'
  tbinfo,sciheader,sci_h
  scitime= tbget(sci_h,scitable,'Time')

  ;; re-interpret 1D table as 2D table
  scitime0=reform(scitime,30,1875)  
  ;; calculate the offsets in seconds
  offset = (metatime - scitime0[0,*]) *3600d *24d 

  ;; print some examples
  print,'quaternion time minus data time in s: '+ $
        'first 10 frames of the day'
  print,offset[0:9],format="(d15.7)"
  print,'quaternion time minus data time in s: '+ $
        'last 10 frames of the day'
  print,offset[1865:1874],format="(d15.7)"
  print,'mean and stand. deviation:'
  print,mean(offset),format="(d15.12)"
  print,sqrt(mean((offset-0.0256)^2)), $
        format="(d15.12)"
end
\end{verbatim}

\end{document}